\documentclass[twocolumn]{aastex6}

\collaborationName{Friends of AASTeX}

\shorttitle{FRB Energetics \& High-$z$ FRBs}
\shortauthors{Zhang}

\begin{document}


\title{FRB Energetics and detectability from high redshifts}


\author{Bing Zhang }
\affil{Department of Physics and Astronomy, University of Nevada, Las Vegas, Las Vegas, NV 89154,
zhang@physics.unlv.edu}
\affil{National Astronomical Observatories, Chinese Academy of Sciences, Beijing 100012, China}

\begin{abstract}
We estimate the upper limit redshifts of known FRBs using the dispersion measure (DM) - redshift ($z$) relation and derive the upper limit peak luminosity $L_p$ and energy $E$ of FRBs within the observational band. The average $z$ upper limits range from 0.17 to 3.10, the average $L_p$ upper limits range from $1.24 \times 10^{42} \rm erg \ s^{-1}$ to $7.80 \times 10^{44} \rm erg \ s^{-1}$, and the average $E$ upper limits range from $6.91 \times 10^{39}$ erg to $1.94 \times 10^{42}$ erg.  FRB 160102 with DM $=2596.1 \pm 0.3 \ {\rm pc \ cm^{-3}}$ likely has a redshift greater than 3. Assuming that its intrinsic DM contribution from the host and FRB source is ${\rm DM_{host}+DM_{scr}}\sim 100 \ {\rm pc \ cm^{-3}}$, such an FRB can be detected up to $z \sim 3.6$ by Parkes and by FAST under ideal conditions up to  $z \sim 10.4$. Assuming that there exist FRBs detectable at $z\sim 15$ by sensitive telescopes such as FAST, the upper limit DM for FRB searches may be set to $\sim 9000 \ {\rm pc \ cm^{-3}}$. For single-dish telescopes, those with a larger aperture tend to detect more FRBs than those with a smaller aperture if the FRB luminosity function index $\alpha_{\rm L}$ is steeper than 2, and vice versa. In any case, large-aperture telescopes such as FAST are more capable of detecting high-$z$ FRBs, even though most of FRBs detected by them are still from relatively low redshifts.
 \end{abstract}

\section{Introduction} \label{sec:intro}

Fast radio bursts (FRBs) \citep{lorimer07,thornton13,petroff15b,petroff16,katz18} are mysterious radio transients with excess dispersion measure (DM) with respect to the Galactic values. The localization of the only repeating source FRB 121102 \citep{spitler16,scholz16,chatterjee17,marcote17,tendulkar17} confirmed the cosmological origin of at least this source (at $z=0.19$), and there is a good reason to believe that most, if not all, FRBs also originate from cosmological distances \citep{thornton13,caleb16,macquart18}. If many FRBs are localized so that their redshifts are measured, the combined $z$ and DM information of these events can be used to directly measure the baryon number density of the universe \citep{deng14,keane16} and its large-scale fluctuation \citep{mcquinn14}, constrain the cosmological parameters together with other cosmic probes \citep{zhou14,gao14,walters18}, constrain the cosmic ionization history \citep{deng14,zheng14,fialkov16}, measure the Hubble Constant and cosmic curvature if some repeating FRBs are  gravitationally lensed  \citep{li18}, and even constrain Einstein's Weak Equivalence Principle (WEP) \citep{wei15} and the rest mass of the photon \citep{wu16b,shao17}. It is not known whether FRBs can be made at high redshifts. Certain progenitor models \citep[e.g.][]{zhang14,connor16,cordes16,metzger17} make connections between FRBs and young neutron stars produced from supernovae or gamma-ray bursts (GRBs), so that their birth rate may track the star formation history of the universe. Since GRBs with redshifts up to 9.4 have been detected \citep[e.g.][]{cucchiara11}, it is possible that some FRBs may be generated at high redshifts within these scenarios. Detecting high-redshift (e.g. $z > 7$) FRBs is extremely valuable, since they can be used to probe the reionization history of the universe and place the most stringent constraints on the WEP and the mass of the photon.

Many current and upcoming facilities have FRB detections as one of their leading scientific goals (e.g. Parkes \citep{petroff16}, UTMOST \citep{UTMOST}, CHIME \citep{chime}, FAST \citep{li18}, ASKAP \citep{johnson09}, MeerKAT \citep{booth09}, SKA \citep{macquart15,fialkov17}. It is interesting to investigate from how high redshifts the FRBs can be detected with these telescopes.

\begin{table*}\label{table:FRBs}
\begin{deluxetable}{ccccccccccc}
        \tablecaption{The observational properties of a sample of FRBs (including ``all events'' in the FRB catalog as of Aug. 15, 2018, http://www.frbcat.org \citep{petroff16}) and their estimated average upper limits of redshift ($z$), isotropic peak luminosity ($L_p$), and isotropic energy ($E$).}
        \tablehead{
                \colhead{FRB Name}  & \colhead{DM} & \colhead{$\rm DM_E$} & \colhead{$z$} &  \colhead{$S_{\nu,p}$} &  \colhead{$t_{\rm obs}$} &  \colhead{$\nu_c$ \tablenotemark{a}} &  \colhead{$L_p$} &  \colhead{$E$}  &  \colhead{telescope } & \colhead{S/N} \\
                \colhead{(yymmdd)} & \colhead{(${\rm pc \ cm^{-3}}$)} & \colhead{(${\rm pc \ cm^{-3}}$)} & & \colhead{(${\rm Jy}$)} & \colhead{(${\rm ms}$)} & \colhead{(${\rm MHz}$)} & \colhead{($10^{43}{\rm erg/s}$)} & \colhead{($10^{40}{\rm erg}$)} & &
        }

        \startdata
FRB010125 & $790 \pm 3$ & 680 & $<$0.76 & 0.3 & 9.4 & 1372.5 & $<$1.16 & $<$6.22 & Parkes & 17 \\
FRB010621\tablenotemark{b} & $745\pm 10$  & 222 & $<$0.26  & 0.41 & 7 & 1374 & $<$0.124 & $<$0.691 & Parkes & 16.3 \\
FRB010724 &  375  & 330.42 &  $<$0.38 & 30 & 5 & 1374 & $<$21.9 & $<$79.3 & Parkes & 23 \\
FRB090625 &  $899.55 \pm 0.01$ & 867.86 & $<$0.97 & 1.14 & 1.92 & 1352 & $<$11.7 & $<$11.4 & Parkes & 30 \\
FRB110220 &  $944.38 \pm 0.05$ & 909.61 & $<$1.01 & 1.3 & 5.6 & 1352 & $<$18.6 & $<$51.8 & Parkes & 49 \\
FRB110523 &  $623.3 \pm 0.06$  & 579.78 &  $<$0.65 & 0.6 & 1.73 & 800 & $<$0.928 & $<$0.972 & GBT & 42 \\
FRB110626 &  $723 \pm 0.3$ & 675.54 &  $<$0.76 & 0.4 & 1.4 & 1352 & $<$1.53 & $<$1.22 & Parkes & 11 \\
FRB110703 &  $1103.6\pm 0.7$  & 1071.27 &  $<$1.19 & 0.5 & 4.3 & 1352 & $<$5.74 & $<$11.3 & Parkes & 16 \\
FRB120127 &  $553.3 \pm 0.3$ & 521.48 &  $<$0.59 & 0.5 & 1.1 & 1352 & $<$1.03 & $<$0.711 & Parkes & 11 \\
FRB121002 &  $1629.18 \pm 0.02$  & 1554.91 & $<$1.75 & 0.43 & 5.44 & 1352 & $<$12.7 & $<$25.1 & Parkes & 16 \\
FRB121102 & $557\pm 2$ & 369 & $<$0.42 & 0.4 & 3 & 1375 & $<$0.370 & $<$0.782 & Arecibo & 14 \\
FRB130626 &  $952.4\pm 0.1$  & 885.53 & $<$0.99 & 0.74 & 1.98 & 1352 & $<$5.39 & $<$5.36 & Parkes & 21 \\
FRB130628 &  $469.88 \pm 0.01$ & 417.3 & $<$0.48 & 1.91 & 0.64 & 1352 & $<$2.38 & $<$1.03 & Parkes & 29 \\
FRB130729 &  $861 \pm 2$  & 830 & $<$0.92  & 0.22 & 15.61 & 1352 & $<$1.34 & $<$10.9 & Parkes &14 \\
FRB131104 &  $779 \pm 1$  & 707.9 & $<$0.79 & 1.12 & 2.08 & 1352 & $<$4.69 & $<$5.45 & Parkes & 30 \\
FRB140514 &  $562.7 \pm 0.6$  & 527.8 & $<$0.60 & 0.471 & 2.8 & 1352 & $<$1.00 & $<$1.76 & Parkes & 16 \\
FRB150215 &  $1105.6 \pm 0.8$ & 678.4 & $<$0.76 & 0.7 & 2.88 & 1352 & $<$2.68 & $<$4.38 & Parkes & 19 \\
FRB150418 &  $776.2 \pm 0.5$  & 587.7 & $<$0.66 & 2.2  & 0.8 & 1352 & $<$5.93 & $<$2.85 & Parkes & 39 \\
FRB150610 &  $1593.9 \pm 0.6$ & 1471.9 & $<$1.65 & 0.7 & 2 & 1352 & $<$17.9 & $<$13.5 & Parkes & 18 \\
FRB150807 &  $266.5 \pm 0.1$  & 229.6 & $<$0.27 & 128 & 0.35 & 1352 & $<$41.7 & $<$11.5 & Parkes & 0\tablenotemark{c}\\
FRB151206 &  $1909.8 \pm 0.6$ & 1749.8 & $<$1.99 & 0.3 & 3 & 1352 & $<$12.1 & $<$12.2 & Parkes & 10 \\
FRB151230 &  $960.4 \pm 0.5$ & 922.4 & $<$1.03 & 0.42 & 4.4 & 1352 & $<$3.36 & $<$7.28 & Parkes & 17 \\
FRB160102 &  $2596.1 \pm 0.3$ & 2583.1 & $<$3.10 & 0.5 & 3.4 & 1352 & $<$59.2 & $<$49.1 & Parkes & 16 \\
FRB160317 &  $1165 \pm 11$  & 845.4 & $<$0.94 & 3 & 21 & 843 & $<$12.0 & $<$129 & UTMOST & 13 \\
FRB160410 &  $278 \pm 3$  & 220.3 & $<$0.26 & 7 & 4 & 843 & $<$1.30 & $<$4.13 & UTMOST & 13 \\
FRB160608 &  $682 \pm 7$  & 443.7 & $<$0.50 & 4.3 & 9 & 843 & $<$3.69 & $<$22.1 & UTMOST & 12 \\
FRB170107 &  $609.5 \pm 0.5$  & 574.5 & $<$0.65 & 22.3 & 2.6 & 1320 & $<$56.9 & $<$89.6 & ASKAP & 16 \\
FRB170827 &  $176.4 \pm 0$ & 139.4 & $<$0.17 & 50.3 & 0.4 & 835 & $<$3.57 & $<$1.22 & UTMOST & 90 \\
FRB170922 & 1111 & 1066 & $<$1.19 & 2.3 & 26 & 835 & $<$16.3 & $<$194 & UTMOST & 22 \\
FRB171209 & 1458 & 1445 & $<$1.62 & 0.92 & 2.5 & 1352 & $<$22.6 & $<$21.5 & Parkes & 40 \\
FRB180301 & 520 & 365 & $<$0.42 & 0.5 & 3 & 1352 & $<$0.455 & $<$0.962 & Parkes & 16 \\
FRB180309 & 263.47 & 218.78 & $<$0.26 & 20.8 & 0.576 & 1352 & $<$6.20 & $<$2.84 & Parkes & 411 \\
FRB180311 & 1575.6 & 1530.4 & $<$1.72 & 0.2 & 12 & 1352 & $<$5.68 & $<$25.1 & Parkes & 11.5 \\
FRB180528 & 899 & 830 & $<$0.92 & 13.8 & 1.3 & 835 & $<$51.7 & $<$35.0 & UTMOST & 14 \\
FRB180714 & 1469.873 & 1212.873 & $<$1.35 & 5 & 1 & 1352 & $<$78.0 & $<$33.2 & Parkes & 20 \\
FRB180725A\tablenotemark{d} & 716.6 & 647.6 & $<$0.73 & & 2 & 600 &  & & CHIME & 20.6 \\
        \enddata
\end{deluxetable}
\tablenotetext{a}{Notice that $\nu_c$ can be different for the same telescope. The values presented are the ones
reported in the original discovery papers.}
\tablenotetext{b}{This FRB reached saturation so that the peak flux and S/N reported \citep{lorimer07} was greatly under-estimated.}
\tablenotetext{c}{No S/N was reported in the original paper \citep{ravi16}. } 
\tablenotetext{d}{No flux was reported in the original ATel \citep{boyle18}.}
\end{table*}

\section{Estimates of $z$ and $E$ of known FRBs}

The observed DM of an FRB can be decomposed into
\begin{equation}
 \rm DM  =   \rm DM_{MW} + DM_E 
\end{equation}
where
\begin{equation}
{\rm DM_E = DM_{IGM}} + \frac{\rm DM_{host} + DM_{src}}{1+z}
\end{equation} 
is the external DM contribution outside the Milky Way galaxy, and $\rm DM_{host}$ and $\rm DM_{src}$ are the DM contributions from the FRB host galaxy and source environment, respectively, in the cosmological rest frame of the FRB. The measured values of both are smaller by a factor of $(1+z)$ \citep{ioka03,deng14}.  The IGM portion of DM is related to the distance (redshift) of the source through \citep{deng14}
\begin{equation}
 {\rm DM_{IGM}}  =  \frac{3c H_0 \Omega_b f_{\rm IGM} }{8 \pi G m_p} \int_0^z \frac{\chi(z) (1+z) dz} {[\Omega_m (1+z)^3+\Omega_\Lambda]^{1/2}}
\label{eq:DM-IGM1}
\end{equation}
in the flat $\Lambda$CDM universe (i.e., the dark energy equation of state parameter $w= -1$), 
where $\Omega_b$ is the baryon density, $H_0$ is Hubble constant, $f_{\rm IGM} \sim 0.83$ is the fraction of baryons in the IGM \citep{fukugita98}\footnote{In principle, $f_{\rm IGM}$ can be redshift-dependent. Here we adopt an average value by assuming that the redshift evolution effect is not significant.},  
\begin{equation}
\chi(z) = \frac{3}{4} y_1 \chi_{e,\rm H}(z) + \frac{1}{8} y_2 \chi_{e,\rm He}(z)
\label{eq:chi}
\end{equation} 
denotes the free electron number per baryon in the universe, with $\chi_{e,\rm H}$ and $\chi_{e,\rm He}$ denoting the ionization fraction of hydrogen and helium, respectively, and $y_1 \sim y_2 \sim 1$ denoting the possible slight deviation from the 3/4 - 1/4 split of hydrogen and helium abundance in the universe. If both hydrogen and helium are fully ionized (valid below $z \sim 3$), one has $\chi(z) \simeq 7/8$. Adopting the latest Planck results \citep{planck} for the $\Lambda$CDM cosmological parameters, i.e. $H_0 = 67.74\pm 0.46 \ {\rm km \ s^{-1} \ kpc^{-1}}$, $\Omega_b = 0.0486 \pm 0.0010$, $\Omega_m = 0.3089 \pm 0.0062$, $\Omega_\Lambda = 0.6911 \pm 0.0062$, Equation (\ref{eq:DM-IGM1}) has the numerical value
\begin{eqnarray}
 {\rm DM_{IGM}} & \simeq & 1112 \ {\rm pc \ cm^{-3}}  f_{\rm IGM} \chi F(z) \nonumber \\
 & \simeq & 807  \ {\rm pc \ cm^{-3}} \frac{f_{\rm IGM}}{0.83} \frac{\chi}{7/8} F(z),
\label{eq:DM-IGM2}
\end{eqnarray}
where 
\begin{equation}
F(z) = \int_0^z \frac{(1+z) dz} {[\Omega_m (1+z)^3+\Omega_\Lambda]^{1/2}},
\label{eq:F}
\end{equation} 
which lies in the range $1 - 1.12$ for $z < 3$. If one adopts an average value $F(z) \sim 1.06$, one has ${\rm DM_{IGM}} \sim 1168 \ {\rm pc \ cm^{-3}}  f_{\rm IGM} \chi z = 855 \ {\rm pc \ cm^{-3}}  (f_{\rm IGM}/0.83) (\chi/(7/8)) z$ for $z < 3$. In the FRB literature,  $z \sim {\rm DM_{E}}/ (1200  \ {\rm pc \ cm^{-3}})$ has been adopted \citep{petroff16,caleb16} to estimate the upper limit of the FRB redshifts based on the earlier calculations by \cite{ioka03} and \cite{inoue04}. These calculations have assumed that essentially all baryons are in the IGM ($f_{\rm IGM} \sim 1$) and that the universe is composed of hydrogen only ($\chi = 1$), which significantly {\em under-estimate} the redshift upper limit $z$ for a given $\rm DM_E$ (by a factor of $\sim 0.83\cdot (7/8) \sim 0.73$).  According to our results, a rough estimate 
\begin{equation}
 z \sim {\rm DM_{IGM} / 855 pc \ cm^{-3}} 
\end{equation}
is recommended for $z<3$, which has a $\sim 6\%$ error. Notice that this relation is valid on average. Due to the existence of large-scale structures, different lines of sights may give different $\rm DM_{IGM}$ values for the same $z$ \citep{mcquinn14}. The variation is redshift-dependent, and can be up to $\sim 40\%$ at $z \sim 1$ and drops at higher redshifts. If one adopts the $\sim 40\%$ variation, the conversion factor 855 would be in the range $\sim (510-1200)$.

In order to derive $\rm DM_{IGM}$ of an FRB, one needs to know $\rm DM_{host} + DM_{src}$. This is difficult to derive from an individual FRB, but may be derived statistically using a sample of FRBs \citep{yangzhang16,yang17}. The observations of FRB 121102 \citep{chatterjee17,marcote17,tendulkar17} and a statistical study \citep{yang17} suggest that this sum is not small, which is comparable to $\rm DM_{IGM}$ for FRB 121102 (if the true $\rm DM_{IGM}$ of the source is close to the average value derived in Eq.(\ref{eq:DM-IGM1})). In any case, $\rm DM_E$ can be used to derive an {\em average} upper limit of $\rm DM_{IGM}$, and hence, an  {\em average} upper limit of $z$, of a particular FRB (again noticing the fluctuations of $\rm DM_{IGM}$ along different lines of sight \citep{mcquinn14}). As DM increases, this average upper limit gets closer to the true value due to the $(1+z)$ suppression factor of $\rm DM_{host}+DM_{scr}$. The average $z$ upper limits of the published FRBs (extracted from the FRB catalog, \citealt{petroff16}) are presented in Table \ref{table:FRBs}. The external $\rm DM_E$ values are directly taken from FRB catalog, which were presented in the original papers that reported the discovery of each FRB \citep[][and references therein]{petroff16}. In those original papers, some authors have used of the Galactic electron density model of NE2001 \citep{cordes02} while some others used YMW17 \citep{yao17}. The $\rm DM_{MW}$ values derived from the two models usually agree with each other, but could be very different for some FRBs. In any case, since $\rm DM_{MW}$ is usually a small portion of the total DM, the derived $\rm DM_E$ from the two Galactic electron density model would not differ significantly, and the conclusions presented in this paper are essentially not affected. In the derivations of $\rm DM_E$ of these original papers, the DM contribution from the Galactic halo \citep[e.g.][]{dolag15} was not deducted.

With the $z$ upper limit, one can derive the upper limit of the {\em isotropic peak luminosity} and {\em isotropic energy} of an FRB within the observed bandwidth, which read
\begin{eqnarray}
L_p & \simeq & {4\pi D^2_{\rm L}} {\cal S}_{\nu,p} \nu_c \nonumber \\
&= & (10^{42} \ {\rm erg \ s^{-1}}) {4\pi} \left(\frac{D_{\rm L}}{10^{28} \ {\rm cm}}\right)^2 \frac{{\cal S}_{\nu,p}}{\rm Jy} \frac{\nu_c}{\rm GHz}, 
\label{eq:Lp} \\
E & \simeq & \frac{4\pi D^2_{\rm L}}{(1+z)} {\cal F_\nu} \nu_c \nonumber \\
&= & (10^{39} \ {\rm erg}) \frac{4\pi}{(1+z)} \left(\frac{D_{\rm L}}{10^{28} \ {\rm cm}}\right)^2 \frac{\cal F_\nu}{\rm Jy \cdot ms} \frac{\nu_c}{\rm GHz},
\label{eq:E}
\end{eqnarray}
where $S_{\nu,p}$ is the specific peak flux (in units of $\rm erg \ s^{-1} \ \ cm^{-2} \ Hz^{-1}$ or Jy), ${\cal F_\nu} = S_{\nu,p} \tau_{\rm obs}$ is the specific fluence (in units of $\rm erg \ cm^{-2} \ Hz^{-1} $, or $\rm Jy \cdot ms$). Notice that Eq.(\ref{eq:E}) is different from the formula used in some previous, influential papers including the FRB catalog paper \citep{petroff16,caleb16} in two aspects. First, we use the central frequency $\nu_c$ rather than the bandwidth $B$ of the telescope to derive $L_p$ and $E$. We believe that this is more appropriate. Bandwidth $B$ makes a connection between the detected energy and fluence, but for estimating the source energy, one should use the central frequency $\nu_c$. Let us consider the same FRB detected by two telescopes with the same $\nu_c$ but different bandwidths $B$. The telescope with a wider band receives more energy than the with a narrower band, but their derived specific flux (energy per unit frequency per unit time per unit area) should be the same. When one estimates the luminosity and energy of the source, the formula of \cite{petroff16,caleb16} would give two different values for the same source, which is apparently incorrect. One may also consider two telescopes with the same $B$ but operating at two different $\nu_c$ values. If these two telescopes each detected an FRB with the same specific flux/fluence, using the formula of \cite{petroff16,caleb16} would give rise to the same $L_p$ and $E$ for the two FRBs, while in reality the burst detected in the higher frequency band should have higher $L_p$ and $E$ than the other one. Therefore using $\nu_c$ to calculate $L_p$ and $E$ is more reasonable. 
Second, the factor $(1+z)$ was applied incorrectly in those papers when connecting specific fluence with the FRB energy\footnote{According to Eq.(2) of \cite{petroff16} and Eq.(2) of \cite{caleb16}, one has $E = 4\pi D_{\rm L}^2 (1+z) {\cal F_\nu} B$, with the $(1+z)$ factor in the numerator rather than in the denominator. }. The definition of luminosity distance $D_{\rm L}$ is such that the luminosity $L$ (in units of $\rm erg \ s^{-1}$) and flux $S$ (in units of $\rm erg \ s^{-1}cm^{-2}$ or $\rm Jy \ Hz$) are connected through $L = 4 \pi D_{\rm L}^2 S$. Multiplying this by the burst-frame intrinsic time $\tau = \tau_{\rm obs}/(1+z)$, one gets energy, which is Eq.(\ref{eq:E}), noticing $S \tau_{\rm obs} = {\cal F} = {\cal F_\nu} \nu_c$,  where ${\cal F}$ is the fluence (in units of $\rm erg \ cm^{-2}$ or $\rm Jy \ ms \ Hz$).

The results are presented in Table \ref{table:FRBs}. Without knowing $\rm DM_{host}$ and $\rm DM_{src}$ and their distributions, one can only present the upper limits of $z$, $L_p$ and $E$. Since there are line-of-sight fluctuations \citep{mcquinn14}, one can only present the average values.

For the FRB sample published in the FRBCAT so far, the average $z$ upper limit ranges from 0.17 (FRB 170827, \citealt{farah17}) to 3.10 (FRB 160102, \citealt{bhandari18}). The average isotropic peak luminosity $L_p$ upper limit ranges from $1.24 \times 10^{42} \ {\rm erg \ s^{-1}}$ (FRB 010621, \citealt{keane12}) to $7.80 \times 10^{44} \ {\rm erg \ s^{-1}}$ (FRB 180714, \citealt{oslowski18}) with a spread of 2.80 dex. The average isotropic energy $E$ upper limit ranges from $6.91 \times 10^{39} \ {\rm erg}$ (FRB 010621) to $1.94 \times 10^{42} \ {\rm erg}$ (FRB 170922, \citealt{farah17b}) with a spread of 2.45 dex.

\section{Detectability of high-$z$ FRBs}

With the Parkes telescope, an FRB with an average redshift upper limit $z \sim 3.10$ was already detected (FRB 160102 with $\rm DM_E \sim 2583 \ pc \ cm^{-3}$). This burst has the second highest average $L_p$ upper limit ($5.69 \times 10^{44} \ {\rm erg \ s^{-1}}$) and has a signal-to-noise ratio (S/N) 16 at Parkes, which means that it may be detectable at an even higher redshift.

To investigate from how high a redshift a particular FRB can be detected, one needs to make the assumptions about $\rm DM_{host}+DM_{src}$ and the spectral shape of the FRB. Observationally the two DM terms are coupled and not easy to differentiate, even though the information of rotation measure may help to break the degeneracy \citep{caleb18}. The host component $\rm DM_{host}$ has been studied based on the observations of different types of galaxies \citep[e.g.][]{xu15,luo18}. The typical value is a few tens $\rm pc \cdot cm^{-3}$. The source component $\rm DM_{src}$ depends on FRB progenitor models and can be large for some models that invoke a dense circumburst medium such as a supernova remnant \citep[e.g.][]{murase16,piro16,metzger17,yangzhang17}. A relatively large value of $\rm DM_{host}+DM_{src}$ was inferred for FRB 121102 \citep{tendulkar17} and from a statistical analysis \citep{yang17}. To balance different considerations, we assume that the intrinsic value of $\rm DM_{host} + DM_{src} \sim 100 pc \cdot cm^{-3}$. The observed value of this sum is smaller by a factor of $(1+z)$.\footnote{For a larger value of  $\rm DM_{host}+DM_{src}$, as suggested by FRB121102, the estimates to $z$, $L_p$ and $E$ for nearby events would be smaller (and more uncertain), but our discussion about the high-$z$ FRBs is not significantly affected due to the $(1+z)$ suppression factor.}

Let us consider an FRB with the observed peak specific flux $S_{\nu,p}$, duration $\tau_{\rm obs}$, and redshift $z$. Now imagine this FRB is moved to a higher redshift $z'$, its peak specific flux $S'_{\nu,p}$ in the same observational frequency band can be calculated as
\begin{eqnarray}
 S'_{\nu,p} & = & \frac{k L_p}{4 \pi (D'_{\rm L})^2 \nu_c} \frac{\hat\tau_{\rm obs}}{\hat\tau'_{\rm obs}}  
= k S_{\nu,p}  \left(\frac{D_{\rm L}}{D'_{\rm L}}\right)^2 \frac{\hat\tau_{\rm obs}}{\hat\tau'_{\rm obs}} ,
\label{eq:Sp}
\end{eqnarray}
where $\hat\tau_{\rm obs} = \tau_{\rm obs}/(1+z)$, $\hat\tau'_{\rm obs} = \tau'_{\rm obs}/(1+z')$ are cosmological-frame equivalence of the observed duration, and
\begin{equation}
 k = \frac{\int_{\nu_a(1+z')}^{\nu_b(1+z')} L_\nu d\nu} {\int_{\nu_a(1+z)}^{\nu_b(1+z)} L_\nu d\nu} = \left(\frac{1+z'}{1+z}\right)^{1-\alpha}
 \label{eq:k}
\end{equation}
is the $k$-correction factor, with $(\nu_a, \nu_b)$ denoting the frequency range of the observational band (with central frequency $\nu_c$). The right-most term of Eq.(\ref{eq:k}) has applied the assumption of a power law FRB spectrum, i.e. $L_\nu \propto \nu^{-\alpha}$.

The observed FRB duration (also called width in the literature) may be written as 
\begin{equation} 
 \tau_{\rm obs} = \left(\tau_{\rm int}^2 (1+z)^2 + \tau_{\rm scat}^2 + \tau_{\rm ins}^2 \right)^{1/2},
\end{equation}
where $\tau_{\rm int}$ is the intrinsic duration of the FRB in the cosmological frame, 
\begin{equation}
\tau_{\rm scat} = \left(\tau_{\rm MW}^2 + \tau_{\rm IGM}^2 + \tau_{\rm host}^2 (1+z)^2\right)^2
\end{equation} 
is the duration due to plasma scattering, which includes contributions from the MW, IGM, and the host (including the host galaxy and the immediate environment of the FRB source), and 
\begin{equation}
\tau_{\rm ins} = \left(\tau_{\rm DM}^2 + \tau_{\rm \delta DM}^2 + \tau_{\delta \nu}^2 + \tau_{\rm samp}^2
\right)^2
\end{equation}
is the instrument-related duration \citep{cordes03,caleb16}, where 
\begin{equation}
\tau_{\rm DM} = 8.3 \mu{\rm s} \ {\rm DM} \Delta \nu_{\rm MHz} \nu_{\rm GHz}^{-3}
\label{eq:tauDM}
\end{equation}
is the frequency-dependent smearing due to dispersion measure, $\tau_{\rm \delta DM} = \tau_{\rm DM} (\delta {\rm DM/DM})$ is the smearing due to the error of DM, $\tau_\nu \sim (\Delta \nu)^{-1} = 1 \mu{\rm s}(\Delta \nu_{\rm MHz})^{-1}$ is the smearing due to the band width, and $\tau_{\rm samp}$ is the sampling time (which is typically $>  50\mu{\rm s}$ for most telescopes but is in any case $< 1$ ms). 
Putting everything together, one can write
\begin{eqnarray}
\hat\tau_{\rm obs} & = & \frac{\tau_{\rm obs}}{1+z} = \left[\tau^2_{\rm int}+\tau^2_{\rm host}  + \right. \nonumber \\
& &\left. 
\frac{\tau_{\rm MW}^2+\tau_{\rm IGM}^2+\tau_{\rm DM}^2+ \tau_{\rm \delta{\rm DM}}^2+\tau_{\rm \delta \nu}^2 + \tau^2_{\rm samp}}{(1+z)^2} \right]^{1/2}.
\label{eq:tau}
\end{eqnarray}

In the following, we argue that for FRBs with $z>2$, $\hat\tau_{\rm obs}$ essentially does not vary when {\em the same FRB} is moved to higher redshifts. Out of the many terms that determine the observed duration (width) $\tau_{\rm obs}$, three terms are likely dominating: the intrinsic duration $\tau_{\rm int}$ as is the case of the repeater \citep{spitler16}, the scattering tail term $\tau_{\rm scat}$ as is the case of the Lorimer burst \citep{lorimer07}, as well as the DM smearing term when either of the first two terms is negligibly small. For the three scattering terms, since FRBs are from high Galactic altitudes, $\rm DM_{MW}$ is negligibly small. Between the contributions from the IGM and host, \cite{xu16} showed that the former is negligibly small for typical turbulent properties of the IGM and that the latter can be the dominant term. The negligible scattering from the IGM is also evident from the fact that there is no clear correlation between the observed width and DM for FRBs. As a result, the dominant terms in Eq.(\ref{eq:tau}) are $\tau_{\rm int}^2$, $\tau_{\rm host}^2$, which do not depend on $z$; and $(\tau_{\rm DM}/(1+z))^2$, which is $\propto ({\rm DM}/(1+z))^2$. At a high redshift, DM is dominated by the IGM term. If one neglects the small corrections due to the change of ionization factors as a function of redshift (i.e. ${\rm DM_{IGM}} \propto F(z)$, Eq.(\ref{eq:F})), the function ${\rm DM_{IGM}}/(1+z) \propto F(z)/(1+z)$ initially rises, reaching a peak around $z \sim 4$, and decays at higher $z$.  In the redshift range from $z=2$ to $z=10$, ${\rm DM_{IGM}}/(1+z)$ is essentially constant within 5\% error. As a result, {\em the DM smearing effect is equivalent to the cosmological time dilation effect}. At even higher redshifts (e.g. $z > 10$), $\rm DM_{IGM}/(1+z)$ steadily declines, so that the DM smearing cannot compensate the $(1+z)$ stretching, and $\hat\tau_{\rm obs}$ starts to slowly decrease with an increasing $z$. Since $\chi_{\rm e,He}$ starts to become less than 1 at $z > 3$ \citep{zheng14} and $\chi_{\rm e,H}$ starts to become less than 1 at $z > 6$ \citep{fan06}, this effect is further enhanced if a precise treatment of ionization is conducted.

Finally, in principle there could be a ``tip-of-iceberg'' effect similar to other transients such as GRBs \citep[e.g.][]{lv14}, i.e. the same burst would be detected to have a longer duration if it is detected with a more sensitive telescope, since more emission is observed above the background noise. In principle, $\hat \tau_{\rm obs}$ may be shorter than its true value at a higher redshift, since the S/N drops when $z$ increases. However, for rapidly variable transients such as FRBs, both rising and decaying slopes are very steep so that this effect may be negligible.

Taking $\hat\tau_{\rm obs} \simeq \hat\tau'_{\rm obs}$ and combining Eqs.(\ref{eq:Sp}) and (\ref{eq:k}), one finally gets
\begin{equation}
 S'_{\nu,p} \simeq S_{\nu,p} k  \left(\frac{D_{\rm L}}{D'_{\rm L}}\right)^2 \simeq S_{\nu,p}  \left(\frac{D_{\rm L}}{D'_{\rm L}}\right)^2 \left(\frac{1+z'}{1+z}\right)^{1-\alpha}.
 \label{eq:Snup2}
\end{equation}
One can see that there are two effects to directly reduce the peak flux of an FRB as it is moved to a higher redshift: the increase of the luminosity distance, and the negative $k$-correction (i.e. one is looking at an intrinsically higher frequency in the source frame where the flux is lower due to the power law spectrum given the same observational frequency). The latter applies to the majority of FRBs, but if the spectral slope of an FRB is positive (e.g. some bursts from the repeater \citealt{spitler16,scholz16,law17}), $k$-correction can be actually positive. 

With the above preparation, one may discuss from how far away the current FRBs can be detected. We consider two FRBs with the highest $L_p$ upper limits: FRB 160102 with $\rm DM_E = 2583.1 pc \ cm^{-3}$ (\citealt{caleb18}, $L_p$ upper limit $5.92 \times 10^{44} \rm erg \ s^{-1}$) and FRB 180714 with $\rm DM_E = 1212.873 pc \ cm^{-3}$ (\citealt{petroff16}, $L_p$ upper limit $7.80 \times 10^{44} \rm erg \ s^{-1}$). Both were detected by Parkes, with the signal-to-noise ratio (S/N) 16 and 20, respectively. The spectral indices of FRBs are poorly constrained. We take a typical value $\alpha \sim 1.6$ for radio pulsars \citep[e.g.][]{lorimer95,xilouris96,jankowski18}, which is also consistent with the theoretical prediction of coherent curvature radiation by bunches \citep{yangzhang17b}. Taking S/N=10 as the threshold for detection and assuming $\rm DM_{host} + DM_{src} \sim 100 pc \cdot cm^{-3}$ for both events, one can perform the following estimates\footnote{Here $\rm DM_{IGM}$ is precisely adopted as the average value, and full ionization of He and H have been assumed. In reality, line-of-sight variations of $\rm DM_{IGM}$ would introduce a large error to render the estimated numbers less precise. Other factors, such as the unknown $\rm DM_{host} + DM_{src}$ value and the source direction from the telescope observing beam, would introduce further uncertainties in the estimates.}: FRB 160102 is at $z\sim 3.06$ with $\rm DM_{IGM} \sim 2556 \ pc \ cm^{-3}$ and $L_p \sim 5.74 \times 10^{44} \ {\rm erg \ s^{-1}}$.  To reduce S/N from 16 to 10, the burst can be detected by Parkes up to $z \sim 3.61$ with $\rm DM_{IGM} \sim 2934 \ pc \ cm^{-3}$ and a total observed $\rm DM \sim 2947  \ pc \ cm^{-3}$. FRB 180714 is at $z\sim 1.30$ with $\rm DM_{IGM} \sim 1170 \ pc \ cm^{-3}$ and $L_p \sim 7.12 \times 10^{44} \ {\rm erg \ s^{-1}}$.  To reduce S/N from 20 to 10, the burst can be detected by Parkes up to $z \sim 1.66$ with $\rm DM_{IGM} \sim 1477 \ pc \ cm^{-3}$ and total observed $\rm DM \sim 1877  \ pc \ cm^{-3}$. FRB 160102 has a lower peak luminosity but is detected at a higher redshift than FRB 180714. This is probably because it was detected at a more favorable beam angle.

Telescopes with larger apertures (and hence, higher sensitivities), e.g. the 300-m Arecibo Radio Telescope and the Five-hundred-meter Aperture Spherical Telescope (FAST), will have a better chance to detect FRBs at even higher redshifts. By design, FAST has an effective area $A_{\rm eff} = 50,000 \ {\rm m^2}$ and system temperature $T_{\rm sys} = 25$ K \citep{lidi18}. Compared with the effective area $A_{\rm eff} = 0.6 \pi (64/2)^2 = 1930 \ {\rm m^2}$ and system temperature $T_{\rm sys} = 24$ K \citep{staveley-smith96}, the sensitivity of FAST (characterized by $A_{\rm eff} / T_{\rm sys}$) is about 25 times of that of Parkes. To be more conservative, in the following, we perform the estimate by assuming that FAST is 20 times more sensitive than Parkes.
Again consider FRB 160102. FAST would have detected it with a S/N $\sim 320$. To reduce S/N from 320 to 10, the burst can be detected at $z \sim 10.4$ with $\rm DM_{IGM} \sim 6487 \ pc \ cm^{-3}$ and a total observed $\rm DM \sim 6500  \ pc \ cm^{-3}$ for $\alpha = 1.6$. Here we have not considered the fact that both He and H are partially ionized at such a high redshift, so that the estimated free electron column density, and therefore $\rm DM_{IGM}$, is an upper limit. 

If there exist FRBs at even higher redshifts with even higher luminosities, large telescopes such as FAST may be still able to barely detect them. It would be interesting to estimate the DM value of these events to optimize the search strategy. For $z \sim 15$, the IGM DM value according to Eq.(\ref{eq:DM-IGM1}) sets an upper limit $\rm DM_{IGM} < 8295  \ pc \ cm^{-3}$. Since at such a high redshift, IGM is nearly neutral, the real $\rm DM_{IGM}$ should be much smaller. Even if one assigns a large $\rm DM_{MW} \sim 1000  \ pc \ cm^{-3}$ to reflect their possible low Galactic latitudes, the maximum observed DM may be close to but not exceeding $9000  \rm \ pc \ cm^{-3}$ (the contributions from the host and source is greatly reduced due to the large reduction by a factor $(1+z) \sim 16$, and $\rm DM_{IGM}$ is much smaller than what Eq.(\ref{eq:DM-IGM1}) presents, since $\chi_{e,\rm H}$ and $\chi_{e,\rm He}$ are less than unity at such high redshifts, see also \citealt{fialkov16}). As a result, the upper limit DM for FAST FRB search may be set to\footnote{If FRBs with $\rm DM > 9000 \ pc \ cm^{-3}$ are indeed detected by any current radio telescope, they should have a huge DM contribution from the host/source  (say, $\rm DM_{host}+DM_{src} > 8000 \ pc \ cm^{-3}$) but at a very low redshift (say, $z<0.5$).} $9000 \rm  \ pc \ cm^{-3}$.

For radio telescopes, the collecting area $A$ and the beam solid angle $\Delta \Omega$ satisfies $A \cdot \Delta \Omega \sim$ const. In an Euclidean geometry with an isotropic distribution of the sources, the horizon distance scales as $D_{h} \propto S_{\rm th}^{-1/2} \propto A^{1/2}$ (where $S_{\rm th}$ is the threshold flux above which the source is detectable). Assuming a uniform source event rate density $\dot n$ (number per unit time per unit volume) for a certain type of transient, the total detection rate (number per unit time) scales as $\dot N \propto \dot n V_{\rm horizon} \Delta\Omega \propto \dot n D_h^3 \Delta\Omega \propto \dot n A^{3/2} A^{-1} \propto \dot n A^{1/2}$. For a constant $\dot n$, the detection rate would scale up with an increasing telescope aperture. For cosmological sources such as FRBs, on the other hand, $D_{h}$ should be replaced by $D_{\rm L,h}$, which still satisfies $D_{\rm L,h} \propto S_{\rm th}^{-1/2} \propto A^{1/2}$. However, the horizon volume increases much more slowly than $D_{\rm L}^3$. In general, the detected event rate by a telescope can be written as
\begin{equation} 
 \dot N =\Delta \Omega \int_0^{z_h} dz \frac{dV(z)}{dz} \frac{\dot n_{\rm FRB}(z)}{1+z} \int_{L_{\rm th}(z)}^{L_{\rm max}} \phi(L') dL',
\end{equation}
where $\phi(L') dL' \propto {L'}^{-\alpha_{\rm L}} dL'$ is the FRB luminosity function\footnote{The luminosity function discussed here refers to that in an observational band, which is observational tractable. The bolometric luminosity function may be more intrinsic, but observationally it is difficult to constrain. The power law index of the bolometric luminosity function would be related to $\alpha_{\rm L}$ through the spectral index $\alpha$ as well as the relationship between the bolometric luminosity and the peak frequency of FRBs.},
\begin{equation}
 \frac{dV}{dz} = \frac{c}{H_0} \frac{D_{\rm L}^2(z)}{(1+z)^2 \sqrt{\Omega_m(1+z)^3+\Omega_\Lambda}}
\end{equation}
is the redshift-dependent cosmological volume in $\Lambda$CDM cosmology, and $\dot n_{\rm FRB}(z)$ is the event rate density at redshift $z$. At large redshifts, the horizon volume increase is negligible, so that the increase of $\dot N$ is mostly due to the increase of $\int_{L_{\rm th}(z)}^{L_{\rm max}} \phi(L') dL'$. Given a same $z$ (and $D_{\rm L}$), one has $L_{\rm th} \propto S_{\rm th} \propto A^{-1}$, so that $\dot N \propto \Delta\Omega \int_{L_{\rm th}}^{L_{\rm max}} \phi(L') dL' \propto  \Delta\Omega L_{\rm th}^{1-\alpha_{\rm L}} \propto A^{\alpha_{\rm L}-2}$. The luminosity function of FRBs is poorly constrained with the current data \citep[e.g.][]{caleb16,li17}. If one adopts $\alpha_{\rm L} \sim 2$, a typical value for cosmological transients \citep{sun15}, the dependence on $A$ disappears. As a result, large telescopes such as FAST may have a  detection rate comparable to smaller telescopes such as Parkes. More generally, large-aperture telescopes tend to detect more FRBs if $\alpha_{\rm L}$ is steeper than 2, and vice versa. In any case, the majority of FRBs detected by larger-aperture radio telescopes should be still nearby low-luminosity ones. Only a small fraction may be high-$z$ FRBs not detectable by smaller telescopes.

\section{Summary}

In view that at least FRB 121102 is cosmological and that many more FRBs will be detected with current and upcoming radio telescopes, here we study the cosmological aspect of FRBs, with the focus on the FRB energetics and the prospects of detecting high-$z$ FRBs. Our main conclusions can be summarized as follows.
\begin{itemize} 
\item Adopting a more precise ${\rm DM_{IGM}}-z$ relation \citep{deng14}, the estimated average redshift upper limit of an FRB for a given $\rm DM_E$ is  higher (Table \ref{table:FRBs}). A more precise estimate of the average $z$ upper limit, i.e. $z \sim {\rm DM_E}/855 \ {\rm pc \ cm^{-3}}$ instead of $z \sim {\rm DM_E}/1200 \ {\rm pc \ cm^{-3}}$, is recommended. Since there are line-of-sight fluctuations due to large scale structures \citep{mcquinn14}, the upper limit redshift falls in the range from ${\rm DM_E}/510 \ {\rm pc \ cm^{-3}}$ to ${\rm DM_E}/1200 \ {\rm pc \ cm^{-3}}$.
\item The isotropic peak luminosity and energy in the observed band of an FRB can be estimated using Eqs.(\ref{eq:Lp}) and (\ref{eq:E}). The $(1+z)$ factor was mis-used in the expression of $E$ in some previous papers. The central frequency $\nu_c$ rather than the band width $B$ should be used in these calculations.
\item Considering various terms contributing to the observed duration (width) of the FRB pulses, one can draw the conclusion that the cosmological rest-frame equivalent duration $\hat\tau_{\rm obs} = \tau_{\rm obs} / (1+z)$ is essentially constant regardless whether the duration is dominated by intrinsic duration, host galaxy/source scattering, or DM smearing. The DM smearing effect is comparable to the time dilation effect.
\item One may estimate the peak flux of a pseudo FRB using Eq.(\ref{eq:Snup2}) when a known FRB is moved to a higher redshift. 
\item In the current sample, FRB 160102 with DM $=2596.1 \pm 0.3 \ {\rm pc \ cm^{-3}}$ likely has the highest redshift. Assuming  ${\rm DM_{host}+DM_{scr}} \sim 100 \ {\rm pc \ cm^{-3}}$, this FRB has a peak luminosity $L_p \sim 5.74 \times 10^{44} \ {\rm erg \ s^{-1}}$, which can be in principle detected up to $z \sim 3.61$ by Parkes with an observed DM $\sim 2947 \ {\rm pc \ cm^{-3}}$, and by FAST under ideal conditions up to  $z \sim 10.4$ with an observed DM  $\sim 6500 \ {\rm pc \ cm^{-3}}$.
\item Assuming that there exist FRBs detectable up to $z\sim 15$, the upper limit DM for FRB searches may be set to $\sim 9000 \ {\rm pc \ cm^{-3}}$ for sensitive radio telescopes such as FAST. 
\item For single-dish telescopes, those with a larger aperture tend to detect more FRBs than those with a smaller aperture if the FRB luminosity function index $\alpha_{\rm L}$ is steeper than 2, and vice versa. Even though small telescope arrays (e.g. CHIME, ASKAP, MeerKAT) will detect and localize many more FRBs, large-aperture telescopes such as FAST are more capable of detecting high-$z$ FRBs.
\end{itemize}

Finally, we'd like to stress that detecting high-$z$ FRBs with large-aperture telescopes is very important scientifically. If these sources can be localized so that a secure redshift $z$ is measured, $\rm DM_E$ can be applied to perform unique studies. At high redshifts, $\rm DM_E \sim \rm DM_{IGM}$ and $\rm DM_{IGM}$ fluctuation is significantly reduced. One can then investigate how much $\chi_{e,\rm H}$ and $\chi_{e,\rm He}$ deviate from unity in Eq.(\ref{eq:chi}), so that the state of re-ionization in the IGM can be probed directly. 

\acknowledgments 
The author acknowledges Wei-Wei Zhu and Di Li for asking about the prospects of detection and the search strategy of FRBs with FAST and for discussing the FAST sensitivity, Emily Petroff for the help with the FRB catalog, and Manisha Caleb, Emily Petroff, Duncan Lorimer, Mathew Bailes, and Sarah Burke-Spolaor for discussions, and an anonymous referee for many helpful suggestions that led to improvements of the presentation of the paper.


\end{document}